\documentclass[12pt, twoside,a4paper, draft]{article}
\usepackage[cp1251]{inputenc}
\usepackage[final]{graphicx}
\newcommand{\CopyName}{O.\ Boyko, \ O.\ Martynyuk, \ V.\ Pivovarchik} % Author's name
\newcommand{\NAME}{O.\ P.\ Boyko, O. \ Martynyuk, V. \ Pivovarchik} %
\newcommand{\Year}{200?} % Publishing year (to fill in by the Editors)
\newcommand{\Volume}{?} % Volume (to fill in by the Editors)
\newcommand{\Number}{?} % Issue (to fill in by the Editors)
\newcommand{\Pages}{?--?} % Pages (to fill in by the Editors)
\newcommand{\rightheadtext}{ ON RECOVERING THE SHAPE OF A QUANTUM TREE}  %please give short version of the title, not exceeding  40 symbols
\usepackage[english,russian,ukrainian]{babel}
\usepackage{ms_we4}
 
\renewcommand{\refname}{\refnam}
\vs
\newcommand{\tit}{On recovering the shape of a quantum tree from the spectrum of the Dirichlet boundary problem} % A title of the paper.
\date{}
\begin{document}
%\hbox to \textwidth{\footnotesize\textsc  .\Volume, \No \Number
%\hfill
%{Matematychni Studii. Vol. 60, No. 2 (2023)}
\vspace{0.3in}
\textup{\scriptsize{}} \vs % ������ ���.
\markboth{{\NAME}}{{\rightheadtext}}
\begin{center} \textsc {\CopyName} \end{center}
\begin{center} \renewcommand{\baselinestretch}{1.3}\bf {\tit} \end{center}

\vspace{20pt plus 0.5pt} {\abstract{ \noindent O. Boyko, O. Martynyuk, V. Pivovarchik\ % Author's name (in English).
%\textit{?}\matref \vspace{3pt} \English % A title of the paper (in English).

Spectral problems are considered generated by the Sturm-Liouville equation on equilateral trees with the  Dirichlet boundary conditions at the   pendant vertices and continuity and Kirchhoff's conditions at the interior vertices.  It is proved that there are no co-spectral (i.e., having the same spectrum of such problem) among equilateral trees of $\leq 8$ vertices. All co-spectral trees of $9$ vertices are presented. % Abstract (in English)

}} \vsk
 %  2020 AMS Mathematics Subject Classification.

 %Keywords of the paper (in English) separated by semicolons
\doi{?}%(to fill in by the Editors)
\renewcommand{\refname}{\refnam}

\vskip10pt

% The text of the paper
\section{Introduction}

It is usual in quantum graph theory to   consider  spectral problems generated by the Sturm-Liouville (Schr\"odinger) equations on  equilateral metric  graph domains with the
Neumann or Dirichlet boundary conditions at the graphs  pendant vertices and standard or in other words generalized Neumann (continuity and Kirchhoff's) conditions at its interior vertices. Here the problem of co-spectrality arises as well as in the classical graph theory.

 It was shown in \cite{vB} that there exist co-spectral  graphs (non-isomorphic graphs with the same spectrum of the Sturm-Liouville problem) in quantum graph theory.  The  example mentioned in \cite{vB} shows two non-isomorphic equilateral metric graphs 
of Fig. 1.   

\begin{figure}[h]
 \begin{center}
    \includegraphics [scale= 0.68 ]{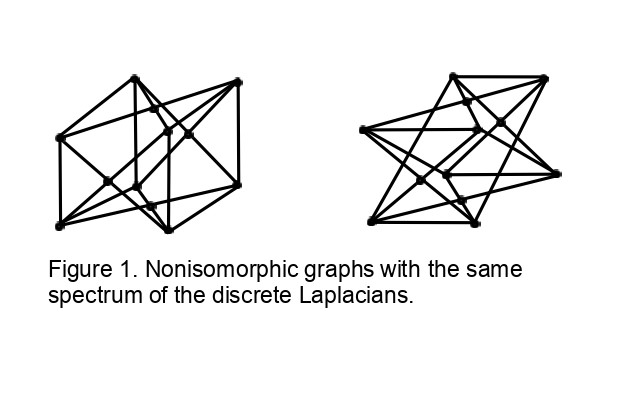}
   \end{center}
\end{figure}

%\end{document}

It should be noticed that in the case of graphs with non-commensurate edges the spectrum uniquely determines the shape of the graph \cite{GS}. 

The spectra of quantum graph problems on equilateral metric graphs are related to the normalized Laplacians of the corresponding combinatorial  graphs in the following way: the eigenvalues of the normalized Laplacian are in one-to-one correspondence with the coefficients in the second term of the asymptotics of the eigenvalues of  the Sturm-Liouville  problem with standard  conditions at the interior vertices and Neumann conditions at the pendant vertices of this graph (see \cite{CP} where the results of \cite{vB1}, \cite{Ca}, \cite{Ex} and of \cite{CaP} were used). This enables to obtain information on the form of a graph using the asymptotics of the eigenvalues.  

Different versions of Ambarzumian's theorem for spectral problems on graphs were proved in  \cite{CP!}, \cite{Kiss}, \cite{Chuan}, \cite{P!}.    
In \cite{KN}, \cite{BKS} a `geometric' Ambarzumian's theorem was proved stating that if the spectrum of the Sturm-Liouville problem with  the Neumann boundary conditions is such as in the case of the problem on a single interval with the zero   potential  then the graph is $P_2$ and  the potential of the Sturm-Liouville equation is zero. In \cite{CP} a geometric Ambarzumian's theorem was proved for connected simple compact equilateral graphs of 5 or less vertices and for trees of 8 or less vertices. This theorem states that if the spectrum of the Sturm-Liouville problem with the Neumann boundary conditions at the pendant vertices and standard conditions at the interior vertices is such as the spectrum of this problem in case of zero potentials on the edges then  this spectrum uniquely determines the shape of the graph and the zero potentials on the edges. 

However, this result cannot be extended to the case of connected simple equilateral graphs of 6 vertices. 

It is known \cite{CP} that  the eigenvalues of the normalized Laplacian can be found from the asymptotics of eigenvalues of the  Sturm-Liouville problem on a graph  not only in the case of `Ambarzumian's' asymptotics. Thus, putting aside Ambarzumian's theorem and the potentials admitting them to  be arbitrary real $L_2$ functions, we put a question: 
can we find the shape of a simple connected equilateral graph using the asymptotics of the eigenvalues of the Sturm-Liouville  spectral problem with the Neumann boundary conditions at the pendant vertices and standard conditions at the interior vertices? It follows from the results of \cite{CP} that if the number of vertices is $\leq 5$ then the spectrum uniquely determines the shape of the (simple connected graph). In case of 6 vertices there is only one pair of co-spectral graphs (see \cite{Pist}, \cite{CP1}. In case of trees the critical number of vertices is 9 (\cite{Pist}, \cite{CP1}). 

In present paper we are interested in what information on
 the shape of a tree can be obtained from the eigenvalue asymptotics in case of the Dirichlet conditions at the pendant vertices. It should be mentioned that in  \cite{B}, \cite{B2} admitting Dirichlet conditions at some of the vertices a method for constructing families of co-spectral systems is proposed, using linear representations of finite groups. Finally, we mention paper \cite{MuP} where the recovering of the shape of a graphs with  an attached lead was considered and where the imaginary part of the Jost function is related with the characteristic function of a spectral problem with the Dirichlet condition at the vertex incident with the lead.

 In Section 2 we formulate the spectral  Sturm-Liouville problem on an  equilateral tree with the standard conditions at the interior vertices and the Dirichlet conditions at the pendant vertices. 
 
 In Section 3 we give some auxiliary results.  We show the relations between the spectrum of the Sturm-Liouville problem and the spectrum of the modified normalized Laplacian of the graphs  interior sub-graph in the case of the Dirichlet conditions at the pendant vertices.

In Section 4 we show that if the number of vertices in a tree does not exceed 8 then the spectrum of the Dirichlet problem uniquely determines the shape of the tree.

In Section 5 we show all pairs or triples of non-isomorphic trees of 9 vertices having the same first  and second terms in the asymptotics of the Dirichlet spectral problem.

\section{Statement of the problems}

Let $G$ be a simple connected  equilateral graph with $p\geq 3$ vertices, $p_{pen}$ pendant vertices, $g$ edges of the length $l$ each. We denote by $v_i$ the vertices, by $d(v_i)$ their degrees, by  $e_j$ the edges.  
We direct each peripheral (incident with the a pendant vertex) edge away from its pendant vertex. 

 Orientation of the rest of the edges is arbitrary.
Thus,  for any interior vertex we consider the indegree  $d^+(v_i)$
and its outdegree by $d^-(v_i)=d(v_i)- d^+(v_i)$. Denote by $W^-(v_i)$ the set of indices $j_s$ ($s=1,2,...,d^-(v_i)$) of the edges outgoing from $v_i$ and by $W^+(v_i)$ the set of indices $k_s$ ($s=1,2,...,d^+(v_i)$) of the edges incoming into $v_i$.

 Local
coordinates for the edges identify each  edge $e_j$ with the interval
$[0,l]$ so that the local coordinate increases in the direction of the edge. This means that each pendant vertex has the
local coordinate $0$. 
Each
interior vertex  has the local coordinate  $l$ on its incoming edge, while the local coordinate of the vertex is $0$ on each outgoing edge. Functions $y_j$ on the edges are
subject to the system of $g$ scalar Sturm-Liouville equations
\begin{equation}
-y_j^{\prime\prime}+q_j(x)y_j=\lambda y_j, \ \ (j=1, 2, ..., g)
\label{2.1}
\end{equation}
where $q_j$ are real-valued functions which belong to
$L_2(0,l)$.  
For each interior vertex  with outgoing edges $e_j$ ($j\in W^-(v_i)$) and
incoming edges $e_k$ ($k\in W^+(v_i)$) the continuity conditions are
\begin{equation}
y_j(0)=y_k(l),  \label{2.2}
\end{equation}
 and  Kirchhoff's condition is
 \begin{equation}
\mathop{\sum}\limits_{k\in W^+(v_i)} y_k^{\prime}(l)=\mathop{\sum}\limits_{j\in W^-(v_i)} y_j^{\prime}(0).
 \label{2.3}
 \end{equation}
We impose the Dirichlet boundary conditions
\begin{equation}
\label{2.4}
y_j(0)=0
\end{equation}
at $r\leq p_{pen}$ of  the pendant vertices and the Neumann conditions
\begin{equation}
\label{2.5}
y_j'(0)=0
\end{equation}
at the rest $p_{pen}-r$ pendant vertices.

Let us denote by $s_j(\sqrt{\lambda},x)$ the solution of the
Sturm-Liouville equation (\ref{2.1}) on the edge $e_j$ which
satisfies the conditions
$s_j(\sqrt{\lambda},0)=s_j^{\prime}(\sqrt{\lambda},0)-1=0$ and by
$c_j(\sqrt{\lambda},x)$ the solution   which satisfies the conditions
$c_j(\sqrt{\lambda},0)-1=c_j^{\prime}(\sqrt{\lambda},0)=0$. 

We look for the solution of problem (\ref{2.1})--(\ref{2.5}) in the form
$$
y_j(\lambda,x)=A_js(\sqrt{\lambda},x)+B_jc(\sqrt{\lambda},x)
$$
Substituting this expression into (\ref{2.1})--(\ref{2.5}) we obtain a system of $2g$ linear algebraic equations with $2g$ unknowns. The determinant of this system can be expressed via the values   
$s_j(\sqrt{\lambda},l)$, $s_j^{\prime}(\sqrt{\lambda},l)$
$c_j(\sqrt{\lambda},l)$ and $c_j^{\prime}(\sqrt{\lambda},l)$ 
for $j=1,2,...,g$. 

The nontrivial solutions of this system correspond to the zeros of the determinant $\phi(\lambda)$ of the systems matrix.
We call this determinant the {\it characteristic
function}  of problem (\ref{2.1})--(\ref{2.5}). 

\section{Auxiliary results}

For a simple graph, the matrix
$A=(A_{i,j})_{i,j=1}^p$ where $A_{i,i}=0$ for all $i=1,2,...,p$ and
for $i\not=j$:
\[\quad A_{i,j}=\left\{\begin{array}{c} 1 \  {\rm if } \ v_i \ {\rm and } \ v_j \ {\rm are  \ adjacent,} \\  0 \ {\rm otherwise}, \end{array}\right.\] 
is called the \textit{adjacency} matrix\index{adjacency matrix}.
Let
$$
D=diag\{d(v_1), d(v_2), ..., d(v_p)\}
$$
be the degree matrix. Then 
$$
\widetilde{ A}=D^{-1/2}AD^{-1/2}
$$
is called the \textit{weighted adjacency} matrix or \textit{normalized Laplacian}.

Let $G$ be a simple connected equilateral graph with $g\geq 2$ edges, $p$ vertices, $p_{pen}$ pendant vertices. Let $r$ ($0\leq r \leq p_{pen}$) be the number of pendant vertices with the Dirichlet conditions.  The graph $\hat{G}$ is obtained by removing the pendant vertices with Dirichlet boundary conditions and the edges  incident with them in $G$. For convenience, we denote by $v_{r+1}$, $v_{r+2}$,..., $v_{p}$ the vertices of $\hat{G}$. Let $\hat{A}$  be the adjacency matrix of ${\hat G}$, let
%$\hat{D}=diag \{\hat{d}(v_1), \hat{d}(v_2), ...,\hat{d}(v_{p-r})\}$ and 
$\hat{D_G}=diag \{d(v_{r+1}), d(v_{r+2}), ..., d(v_{p})\}$, where  $d(v_i)$ is the degree of the vertex $v_i$ in $G$. We consider the polynomial $P_{G,\hat{G}}$ defined by 
$$
\label{3.2}
P_{G,\hat{G}}(z):=det(z\hat{D}_G-\hat{A}).
$$
The following theorem was proved in \cite{MP2} (Theorem 6.4.2). 

\begin{theorem} {\sl Let $G$ be a simple connected graph with at least two edges. Assume that all edges have the same length $l$ and the same potentials symmetric with respect to the midpoints of the edges ($q(l-x)=q(x)$). Then the spectrum of problem (\ref{2.1})--(\ref{2.5})  coincides with the set of zeros of  the characteristic function
$$
\phi_D(\lambda)=s^{g-p+r}(\sqrt{\lambda},l)P_{G,\hat{G}}(c(\sqrt{\lambda},l))
$$
where $s(\sqrt{\lambda},x)$ and $c(\sqrt{\lambda},x)$ are the solutions of the Sturm-Liouville equation which satisfies $s(\sqrt{\lambda}, 0)=s'(\sqrt{\lambda}, 0)-1=0$ and $c(\sqrt{\lambda},0)-1=c'(\sqrt{\lambda},0)=0$. } 
\end{theorem}

\begin{corollary} {\sl Let $T$ be an equilateral  tree with at least 2 edges with the length of each edge $l$ and the same potentials symmetric with respect to the midpoints of the edges ($q(l-x)=q(x)$).  Let $r=p_{pen}$, i.e. let the Dirichlet conditions be imposed at all the pendant vertices. Then 
$$
\phi_D(\lambda)=s(\sqrt{\lambda},l)^{-1+p_{pen}}
P_{T,{\hat T}}(c(\sqrt{\lambda},l)).
$$
If the potentials are zero on all edges then 
$$
\tilde{\phi}_D(\lambda)=\left(\frac{\sin\sqrt{\lambda}l}{\sqrt{\lambda}}\right)
^{-1+p_{pen}}P_{T,\hat{T}}(\cos\sqrt{\lambda}l).
$$
}
\end{corollary}

In the following theorem we consider the case of $r=p_{pen}$, i.e. we impose the Dirichlet condition at each pendant vertex.

\begin{theorem}   
{\sl Let $T$ be an equilateral tree. The eigenvalues of problem (\ref{2.1})--(\ref{2.4})  can be presented as the union of subsequences $\{\lambda_k\}_{k=1}^{\infty}=\mathop{\cup}\limits_{i=1}^{2p-p_{pen}-1}\{\lambda_k^{(i)}\}_{k=1}^{\infty}$
with the following asymptotics
\begin{equation}
\label{3.1}
\sqrt{\lambda_k^{(i)}}\mathop{=}\limits_{k\to\infty}\frac{2\pi (k-1)}{l}\pm\frac{1}{l}\arccos \alpha_{i}+O\left(\frac{1}{k}\right) \ \ {\rm for}  \ \  i=1,2,..., p-p_{pen}, \ \ k=1,2,...
\end{equation}
\begin{equation}
\label{3.3}
\sqrt{\lambda_k^{(i)}}\mathop{=}\limits_{k\to\infty}\frac{\pi k}{l}+O\left(\frac{1}{k}\right) \ \  {\rm for} \ \  i=p-p_{pen}+1, ..., p-1, \ \ k=1,2,...
\end{equation}
Here $\alpha_1, \alpha_2, ..., \alpha_{p-p_{pen}}$ are the zeros of  the polynomial $P_{T,\hat{T}}(z)$.}
\end{theorem}

\begin{proof}[Proof]  From Corollary 1 we obtain the following asymptotics for the case of zero potentials:
\begin{equation}
\label{3.4}
\sqrt{\tilde{\lambda}_k^{(i)}}\mathop{=}\limits_{k\to\infty}\frac{2\pi (k-1)}{l}\pm\frac{1}{l}\arccos \alpha_{i} \ \ {\rm for}  \ \  i=1,2,..., p-p_{pen}, \ \ k=1,2,...
\end{equation}
%\begin{equation} 
%\label{3.5}
%\sqrt{\tilde{\lambda}_k^{(i)}}\mathop{=}\limits_{k\to\infty}\frac{2\pi k}{l}
%-
%\frac{1}{l}\arccos \alpha_{i} \ \  {\rm for} \ \  i=p-p_{pen}+1,..., 2(p-p_{pen}),
%\end{equation}
\begin{equation}
\label{3.6}
\sqrt{\tilde{\lambda}_k^{(i)}}\mathop{=}\limits_{k\to\infty}\frac{\pi k}{l}  \ \ {\rm for} \ \  i=p-p_{pen}+1, ..., p-1, \ \ k=1,2,...
\end{equation}
Here  branch  (\ref{3.4}) is generated by the solutions of the equation $\cos(\sqrt{\lambda}l)=\alpha_i$ while branch (\ref{3.6}) by the solutions of equation $(\frac{\sin(\sqrt{\lambda}l)}{\sqrt{\lambda}})^{-1+p_{pen}}=0$.

By Theorem 5.4 in \cite{CaP} we conclude that $|\lambda^{(j)}_k-{\tilde \lambda}^{(j)}_k|\leq C<\infty$ where  ${\tilde \lambda}^{(j)}_k$ are the eigenvalues of problem (\ref{2.1})--(\ref{2.4}) on the same tree with $q_j\equiv 0$ for all $j$ and therefore, presence of the $L_2(0,l)$-potentials does not influence the first and the second terms of the asymptotics and (\ref{3.1})--(\ref{3.3}) are true. \end{proof}

\begin{remark}
The constants $\alpha_i$ satisfy inequality $|\alpha_i|\leq 1$ because otherwise problem (\ref{2.1})--(\ref{2.4}) would have nonreal eigenvalues. However,   the corresponding operator is selfadjoint (see, e.g. \cite{BK}, Theorem 1.4.4). 
\end{remark}
\begin{remark}
If $\alpha_1=-1$ and $\alpha_{p-p_{pen}}=1$ then the corresponding sequences (\ref{3.1}) can be united to obtain:
$$
\sqrt{\lambda_k^{(i)}}\mathop{=}\limits_{k\to\infty}\frac{\pi (k-1)}{l}+O\left(\frac{1}{k}\right), \ \   \ \, \ \ k=1,(p-p_{pen}).
$$

This sequence is `longer' than each of sequences (\ref{3.3}) by one eigenvalue. Existence of such a sequence (sequences) means existence of a pair (pairs) $\alpha_i=0$ and $\alpha_j=0$.
\end{remark}

\section{Inverse problem}

It is clear from Theorem 2  that looking at the first two terms of the eigenvalue asymptotics we can't distinguish two trees only if the numbers of vertices are the same,  the numbers of edges are the same and the sets $\{\alpha_k\}_{k=1}^{p-p_{pen}}$ corresponding to the two trees coincide. The latter means that the characteristic polynomial $P_{T,\tilde{T}}(z)$ corresponding to one of the  trees is  equal to the characteristic polynomial of the other one multiplied by a non-zero constant. 
 Let us consider all the trees of $\leq 8$ vertices and check whether we can distinguish them using the first two terms of the eigenvalue asymptotics.

 The only tree with $p=3$ vertices is the graph $P_3$, the path of 3 vertices. It has $p_{pen}=2$ pendant vertices. The corresponding polynomial $P_{T,\hat{T}}$ we denote by $\phi_{3,2}(z)$ It is clear that $\phi_{3,2}=-2z$.  

There are two non-isomorphic trees of $p= 4$  vertices. They are $P_4$, the path of 4 vertices with $p_{pen}=2$,  and $S_3$ the star of 3 edges and $p_{pen}=3$. The corresponding polynomials are $\phi_{4.2}=4z^2-1$ and $\phi_{4,3}=-3z$. They have different sets of zeros.

There are three non-isomorphic trees with $p=5$: $P_5$, $S_4$ and the graph of Fig. 2.
The corresponding polynomials are 
$\phi_{5,2}(z)=-8z^3+4z$, $\phi_{5,4}(z)=-4z$ and $\phi_{5,3}(z)=6z^2-1$. Their sets of zeros are different.
\begin{figure}[h]
  \begin{center}
    \includegraphics[scale= 0.68 ] {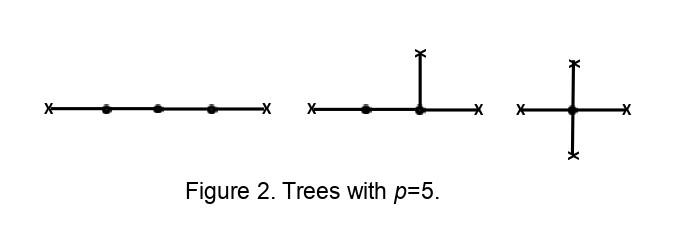}
   \end{center}
\end{figure}

 There is only one tree  $P_6$ with $p=6$, $p_{pen}=2$. The corresponding polynomial is $\phi_{6,2}(z)=16z^4-12z^2+1$. There is  only one tree $S_5$ with $p=6$ and $p_{pen}=5$.  
The corresponding polynomial is $\phi_{6,5}(z)=-5z$.

 There are two  pair of non-isomorphic trees with the same $p$ and $p_{pen}$ among the trees with $p=6$. They are shown at Fig. 3 and Fig. 4.
\begin{figure}[h]
  \begin{center}
    \includegraphics[scale= 0.68] {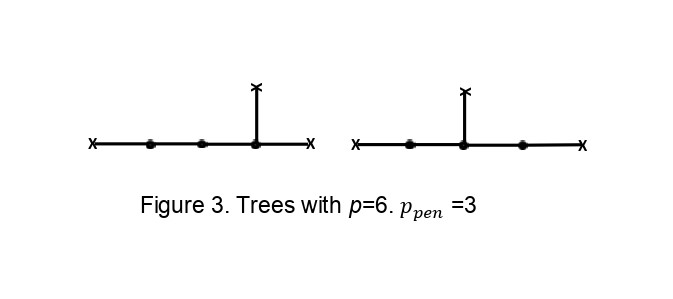}
   \end{center}
\end{figure}
\begin{figure}[h]
  \begin{center}
    \includegraphics[scale= 0.68 ] {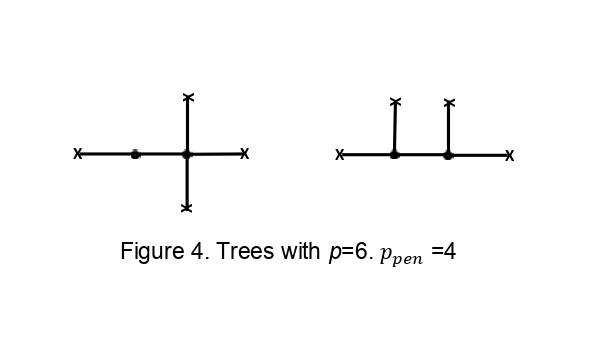}
  \end{center}
\end{figure}

The corresponding polynomials $P_{T,\hat{T}}(z)$ are
 $\phi_{6,3}^1=-12z^3+5z$, $\phi_{6,3}^2=-12z^3+4z$ and $\phi_{6,4}^1=8z^2-1$, $\phi_{6,4}^2=9z^2-1$. Since the sets of zeros of these polynomials are different we conclude that the spectrum of the Dirichlet problem uniquely determines the shape of the tree in case of $p\leq 6$.

 Let $p=7$. Then there is only one tree, namely $P_7$, with $p_{pen}=2$.
The corresponding polynomial is $\phi_{7,2}(z)=-32z^5+32z^3-6z$.
The only  tree with $p=7$ and $p_{pen}=6$ is $S_6$ with the polynomial $\phi_{7,6}=-6z$.

Let $p=7$ and $p_{pen}=3$. There are three such non-isomorphic  trees which are shown at Fig. 5.
\begin{figure}[h]
  \begin{center}
   \includegraphics[scale= 0.68 ] {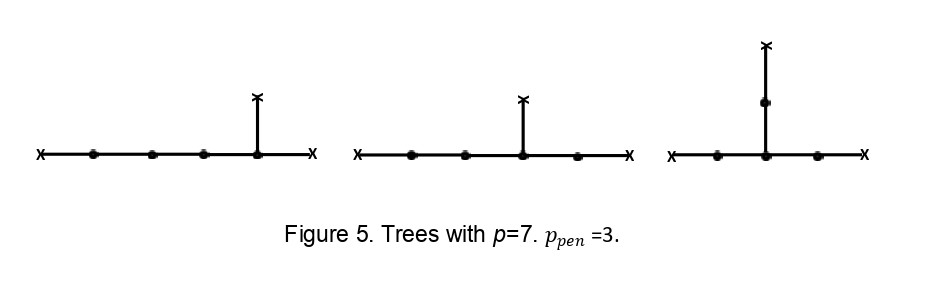}
   \end{center}
\end{figure}

 The corresponding polynomials are 
$$
\phi_{7,3}^1=24z^4-16z^2+1, \  \phi_{7,3}^2=24z^4-14z^2+1, \ \phi_{7,3}^3=24z^4-12z^2.
$$  
The sets of zeros of these polynomials are different and thus the first and the second terms of the eigenvalue asymptotics uniquely determine the shape of the tree in case of $p=7$ and $p_{pen}=3$.
%\end{document}
 Let $p=7$ and $p_{pen}=4$. There are 4 such non-isomorphic  trees which are shown at Fig. 6.  
\begin{figure}[h]
  \begin{center}
    \includegraphics[scale= 0.68 ] {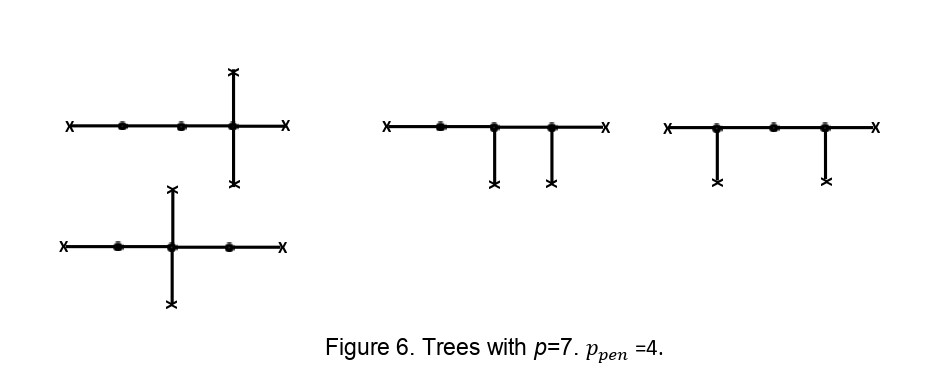}
   \end{center}
\end{figure}
The corresponding polynomials are
$$
\phi_{7,4}^1=-16z^3+6z, \  \phi_{7,4}^2=-18z^3+5z,  \ \phi_{7,4}^3=-18z^3+6z, \ \phi_{7,4}^4=-16z^3+4z.
$$
The sets of zeros of these polynomials are different. Therefore, the first and the second terms of the eigenvalue asymptotics uniquely determine the shape of the tree. 

In case of $p=7$ and $p_{pen}=5$ we face a star and a double star graphs shown at Fig. 7 with the polynomials 
$$ 
\phi_{7,5}^1=10z^2-1, \  \phi_{7,5}^2=12z^2-1.
$$ 
\begin{figure}[h]
  \begin{center}
    \includegraphics[scale= 0.68 ] {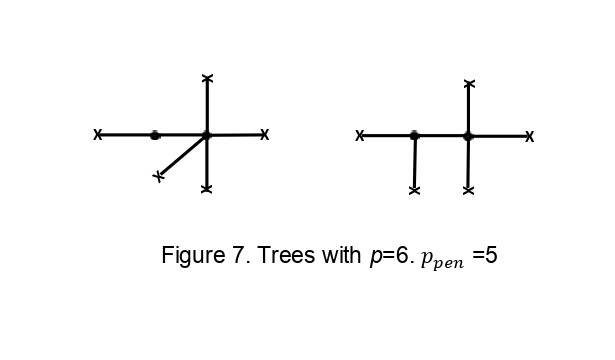}
   \end{center}
\end{figure}

These polynomials have different sets of zeros. Thus, we conclude the first and the second terms of the eigenvalue asymptotics uniquely determine the shape of the tree of 7 vertices.

Now let $p=8$, $p_{pen}=2$. There is only one such tree, namely $P_8$. The corresponding polynomial is $\phi_{8,2}(z)=64z^6-80z^4+24z^2-1$

Also there is only one tree $S_7$ with $p=8$ and $p_{pen}=7$ (the star of 7 edges).The corresponding polynomial is $\phi_{8,7}=-7z$.

Let $p=8$, $p_{pen}=3$. There are 4 such non-isomorphic trees shown at Fig. 8.  
\begin{figure}[h]
  \begin{center}
    \includegraphics[scale= 0.68 ] {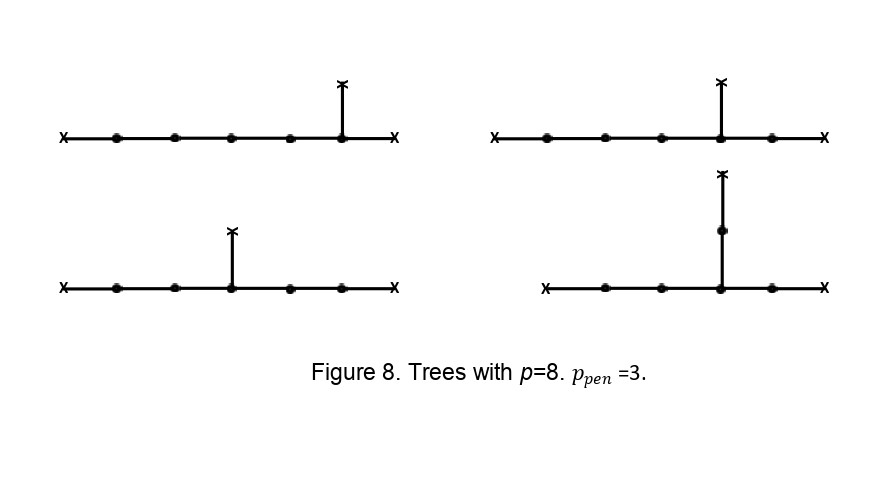}
   \end{center}
\end{figure}

The corresponding polynomials are
$$
\phi_{8,3}^1=-48z^5+44z^3-7z,  \ \phi_{8,3}^2=-48z^5+40z^3-6z, \  \phi_{8,3}^3=-48z^5+40z^3-7z, 
$$
$$
\phi_{8,3}^4=-48z^5+36z^3-4z.
$$
These polynomials have different sets of zeros. Thus, the first and the second terms of the eigenvalue asymptotics uniquely determine the shape of the tree in this case.
 
Let $p=8$, $p_{pen}=4$. There are 8 such non-isomorphic  trees shown at Fig. 9. 
\begin{figure}[h]
  \begin{center}
    \includegraphics[scale= 0.68 ] {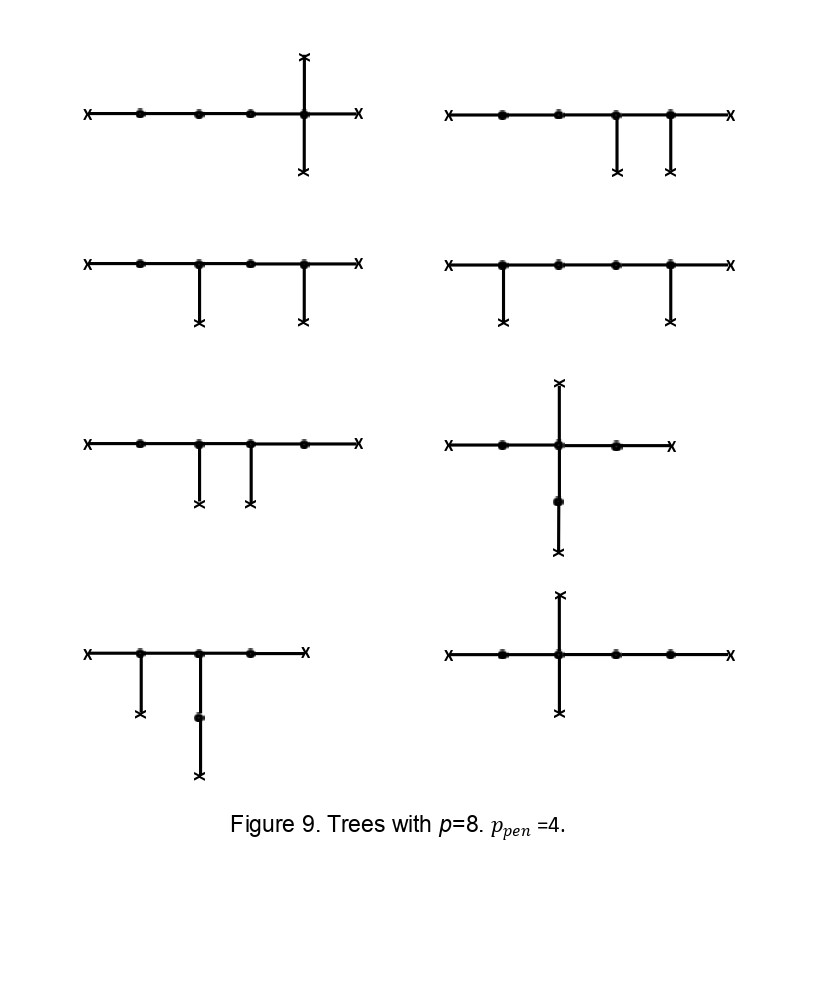}
   \end{center}
\end{figure}

The corresponding polynomials are
$$
\phi_{8,4}^1=32z^4-20z^2+1, \  \phi_{8,4}^2=36z^4-19z^2+1, \  \phi_{8,4}^3=36z^4-18z^2+1, 
$$
$$
 \phi_{8,4}^4=36z^4-21z^2+1, \ \phi_{8,4}^5=36z^4-16z^2+1, \ \phi_{8,4}^6=36z^4-12z^2, 
 $$
 $$
   \phi_{8,4}^7=36z^4-16z^2, \  \phi_{8,4}^8=32z^4-12z^2+1.
$$

Since these polynomials have different sets of zeros, the first and the second terms of the eigenvalue asymptotics uniquely determine the shape of the tree with these data.

Let $p=8$, $p_{pen}=5$. There are 6 such non-isomorphic  trees shown at Fig. 10. 

\begin{figure}[h]
  \begin{center}
    \includegraphics[scale= 0.68 ] {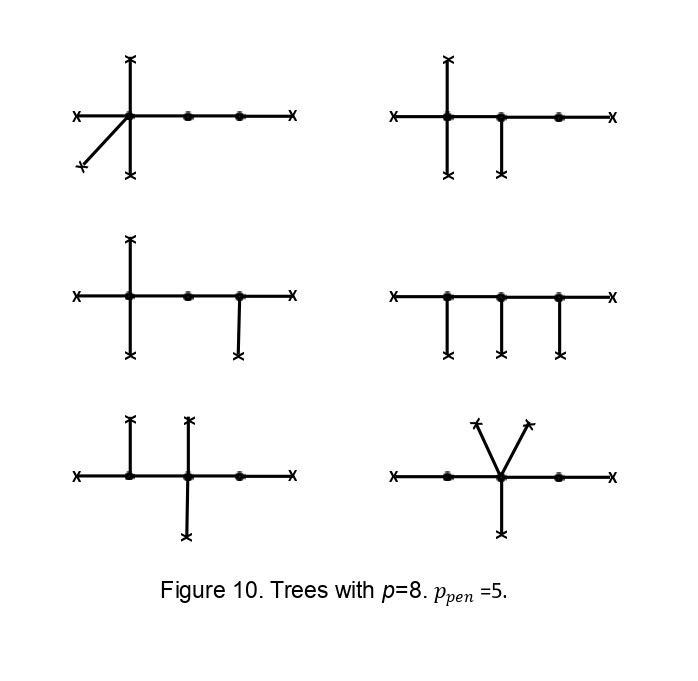}
   \end{center}
\end{figure}

The corresponding polynomials are
$$
\phi_{8,5}^1=-20z^3+7z,  \ \phi_{8,5}^2=-24z^3+6z, \ \phi_{8,5}^3=-24z^3+7z, 
$$
$$
\phi_{8,5}^4=-27z^3+6z, \ \phi_{8,5}^5=-24z^3+5z, \ \phi_{8,5}^6=-20z^3+4z.
$$
All these polynomials have different sets of zeros. Thus, the first and the second terms of the eigenvalue asymptotics uniquely determine the shape of the tree with these data. 

Let $p=8$, $p_{pen}=6$. There are 3 such trees (a star and two double stars) shown at Fig. 11. 
\begin{figure}[h]
  \begin{center}
    \includegraphics[scale= 0.68 ] {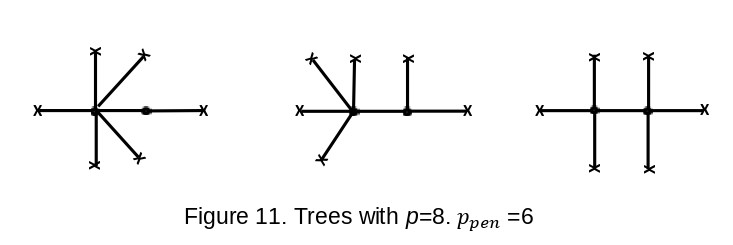}
   \end{center}
\end{figure}

The corresponding polynomials 
$$
\phi_{8,6}^1=12z^2-1,  \phi_{8,6}^2=15z^2-1,  \phi_{8,6}^3=16z^2-1.
$$
have different sets of zeros.

Thus, we have proved the following 

\begin{theorem} {\sl Let $\{\lambda_k\}_{k=1}^{\infty}$ be the spectrum of the Dirichlet spectral problem (\ref{2.1})--(\ref{2.4}) on a simple connected graph. Let $\{\lambda_k\}_{k=1}^{\infty}=\mathop{\cup}\limits_{i=1}^{p-1}\{\lambda_k^{(i)}\}_{k=1}^{\infty}$ where 
\begin{equation}
\label{4.1}
\sqrt{\lambda_k^{(i)}}\mathop{=}\limits_{k\to\infty}\frac{2\pi (k-1)}{l}\pm\gamma_i+O\left(\frac{1}{k}\right) \ \ {\rm for}  \ \  i=1,2,..., \tilde{p}, \ \ k=1,2,...
\end{equation}
\begin{equation}
\label{4.3}
\sqrt{\lambda_k^{(i)}}\mathop{=}\limits_{k\to\infty}\frac{\pi k}{l}+O\left(\frac{1}{k}\right) \ \  {\rm for} \ \  i=\tilde{p}+1, ..., p-1, \ \ k=1,2,...
\end{equation}
and $0<\tilde{p}\leq 6$, $p\leq 8$, $p-\tilde{p}\geq 2$.

Then these asymptotics uniquely determine the shape of the graph as a tree of $\tilde{p}+\tilde{p}_{pen}$ vertices, $\tilde{p}_{pen}$ pendant vertices and $\cos(\gamma_1l), \cos(\gamma_2l), ..., \cos(\gamma_{\tilde{p}}l)$ are the zeros of one of the polynomial $\phi_{i,j}(z)$ or $\phi_{i,j}^s(x)$ described above.}
\end{theorem}

\section{Co-spectral non-isomorphic equilateral trees of 9 vertices}

In this section we consider all  trees of 9 edges and show that there are graphs with the same first and second terms of the asymptotics of eigenvalues of problem (\ref{2.1})--(\ref{2.4}).
These trees are co-spectral in case of zero potentials on all edges.

There is only one tree ($P_9$) with $p=9$ and $p_{pen}=2$ and only one tree of $p=9$ and $p_{pen}=8$ (the star $S_8$). 

Let $p=9$ and $p_{pen}=3$. There are 5 such non-isomorphic trees shown at Fig. 12. 

\begin{figure}[h]
  \begin{center}
    \includegraphics[scale= 0.68 ] {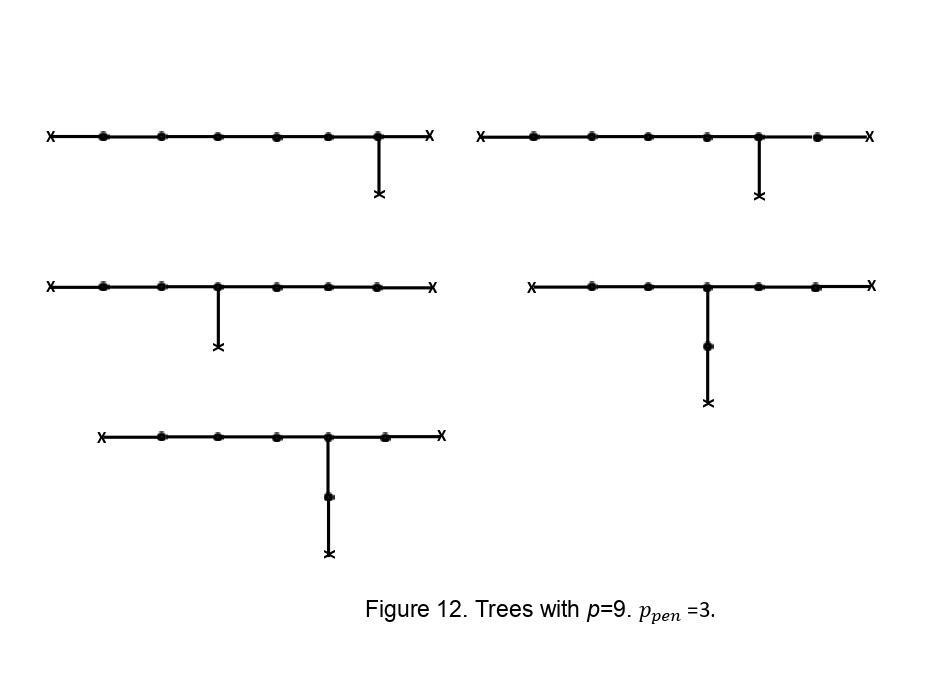}
   \end{center}
\end{figure}

The corresponding  polynomials are
$$
\phi_{9,3}^1=96z^6-112z^4+30z^2-1,  \  \phi_{9,3}^2=96z^6-104z^4+26z^2-1, 
$$
$$
 \phi_{9,3}^3=96z^6-104z^4+28z^2-1, \  \phi_{9,3}^4=96z^6-96z^4+22z^2-1, 
 $$
 $$
 \phi_{9,3}^5=96z^6-96z^4+20z^2.
$$
All these polynomials have different sets of zeros. Thus, we conclude that there are  no  trees in Fig. 12 having the same second terms of eigenvalue asymptotics of the Dirichlet problem. 

Let $p=9$ and $p_{pen}=4$. There are 14 such non-isomorphic trees shown at Fig. 13. 
\begin{figure}[h]
  \begin{center}
    \includegraphics[scale= 0.68 ] {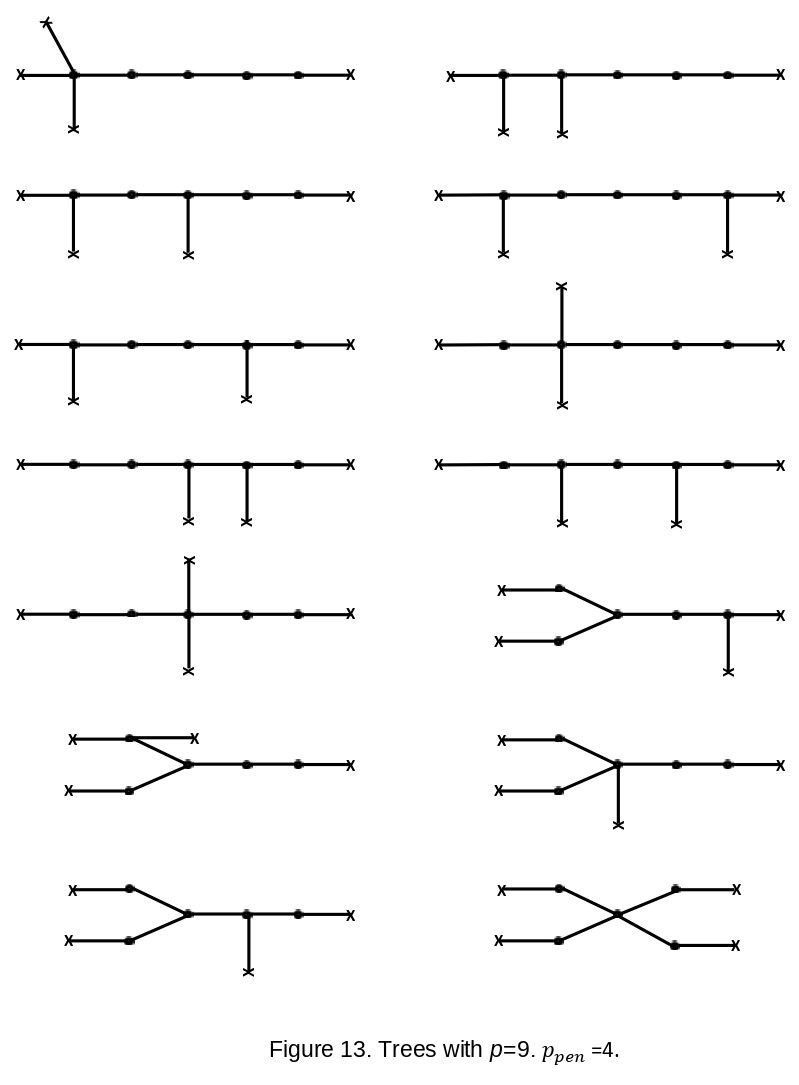}
   \end{center}
\end{figure}
The corresponding polynomials are   
$$
\phi_{9,4}^1=-64z^5+56z^3-8z,  \  \phi_{9,4}^2=-72z^5+56z^3-7z,  \ \phi_{9,4}^3=-72z^5+54z^3-8z, 
$$
$$
\phi_{9,4}^4=-72z^5+60z^3-8z, \ \phi_{9,4}^5=-72z^5+54^3-7z,
$$
$$
\phi_{9,4}^6=-64z^5+48z^3-6z, \ \phi_{9,4}^7=-72z^5+50^3-7z,
$$
$$
\phi_{9,4}^8=-72z^5+48z^3-6z, \ \phi_{9,4}^9=-64z^5+56^3+8z^2-10z-2,
$$
$$
\phi_{9,4}^{10}=-72z^5+48z^3-4z, \  \phi_{9,4}^{11}=-72z^5+50^3-5z,
$$
$$
\phi_{9,4}^{12}=-64z^5+40z^3-4z, \ \phi_{9,4}^{13}=-72z^5+44z^3-4z,  \ \phi_{9,4}^{14}=-64z^5+32z^3.
$$
Since the sets of zeros of these polynomials are different.  Thus, we conclude that there are  no  trees in Fig. 13 having the same second terms of eigenvalue asymptotics of the Dirichlet problem. 

Now let $p=9$ and $p_{pen}=5$. There are 14 such non-isomorphic  trees shown at Fig. 14. 
\begin{figure}[h]
  \begin{center}
    \includegraphics[scale= 0.62 ] {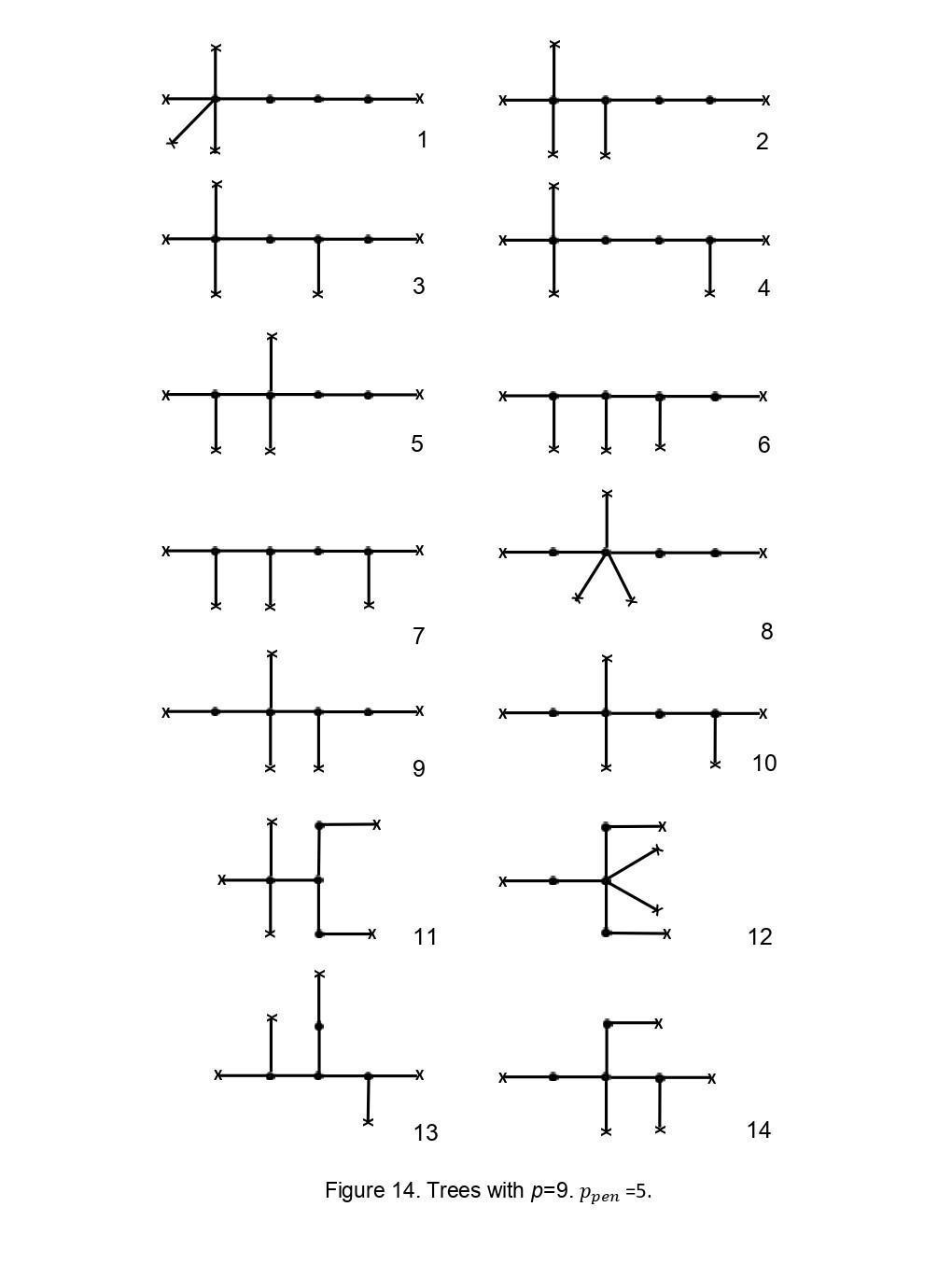}
   \end{center}
\end{figure}

The corresponding  polynomials are
$$
\phi_{9,5}^1=40z^4-24z^2+1, \  \phi_{9,5}^2=48z^4-24z^2+1,
$$ 
$$
  \phi_{9,5}^3=48z^4-22z^2+1,  \ \phi_{9,5}^4=48z^4-26z^2+1,
$$ 
$$
 \phi_{9,5}^5=48z^4-22z^2+1,  \ \phi_{9,5}^6=54z^4-21z^2+1,
$$ 

$$
\phi_{9,5}^7=54z^4-24z^2+1, \ \phi_{9,5}^8=40z^4-18z^2+1,
$$ 

$$
\phi_{9,5}^9=48z^4-18z^2+1, \  \phi_{9,5}^{10}=48z^4-20z^2+1,
$$ 

$$
\phi_{9,5}^{11}=48z^4-20z^2,  \ \phi_{9,5}^{12}=40z^4-12z^2,
$$ 

$$
\phi_{9,5}^{13}=54z^4-21z^2,  \ \phi_{9,5}^{14}=32z^4-12z^2.
$$ 
We see that the graphs 3 and 5 have the same polynomials $\phi_{9,5}^3(z)\equiv\phi_{9,5}^5(z)$ and in case of zero potentials the trees are co-spectral.

Now $p=9$ and $p_{pen}=6$. There are 9 such non-isomorphic trees shown at Fig. 15. Their polynomials are
\begin{figure}[ph]
  \begin{center}
    \includegraphics[scale= 0.7] {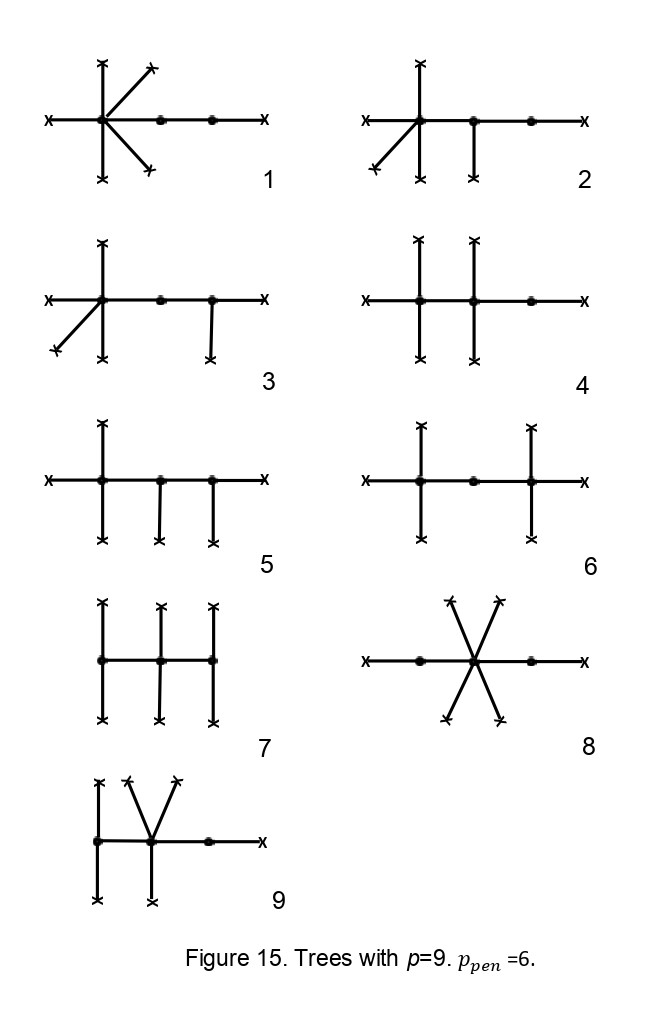}
   \end{center}
\end{figure}

$$
\phi_{9,6}^1=-24z^3+8z, \ \phi_{9,6}^2=-30z^3+7z,
$$ 
$$
\phi_{9,6}^3=-30z^3+8z, \ \phi_{9,6}^4=-32z^3+6z,
$$ 
$$
\phi_{9,6}^5=-36z^3+7z,  \ \phi_{9,6}^6=-32z^3+8z,
$$ 
$$
\phi_{9,6}^7=-36z^3+6z, \  \phi_{9,6}^8=-24z^3+4z, \ \phi_{9,6}^9=-30z^3+5z.
$$ 

%\begin{figure}[h]
%  \begin{center}
%    \includegraphics[scale= 0.88 ] {figDir15}
%   \end{center}
%\end{figure}

The polynomials $\phi_{9,6}^7(z)$, $\phi_{9,6}^8(z)$ and $\phi_{9,6}^9$ have the same set of zeros and consequently in case of zero potentials on all edges the corresponding graphs are co-spectral.

The trees with $p=9$ and $p_{pen}=7$ are double stars shown at Fig.16.
\begin{figure}[h]
  \begin{center}
    \includegraphics[scale= 0.62] {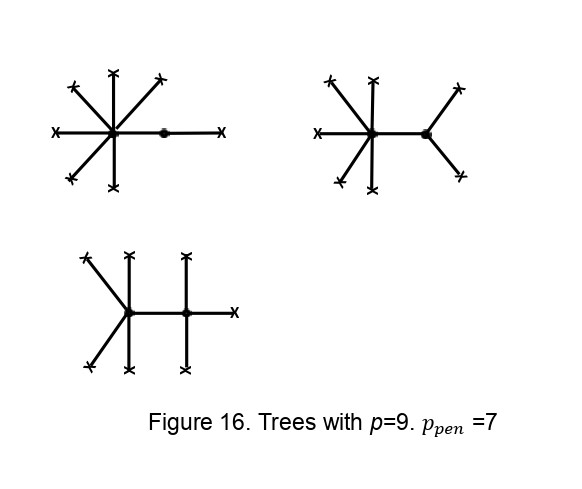}
   \end{center}
\end{figure}

Their polynomials are
$$
P_{T,\hat{T}}=14z^2-1, P_{T,\hat{T}}=18^2-1, P_{T,\hat{T}}=20z^2-1.
$$  
The sets of zeros of these polynomials are different. 

Thus we arrive at 

\begin{theorem}  {\sl Among the trees of 9 verticies there is only one pair (trees 3 and 5 of Fig. 14)  and one triple (trees 7, 8, 9 of Fig. 15) of  equilateral graphs with the same spectrum of the Dirichlet problem with zero potentials on the edges. }
\end{theorem}

\begin{remark}  {\sl In all cases of 9 vertex trees except of trees 3, 5 of Fig.14 and trees 7, 8, 9 of Fig. 15 the first and the second terms of the eigenvalue asymptotics uniquely determine the shape of the tree.}\end{remark}

%\newpage

\noindent\textbf{Acknowledgements}\\

\textit{The authors are grateful to the unknown referee for helpful suggestions}.

\textit{The present research was supported by the Academy of Finland (project no. 358155). The third author is grateful to  University of Vaasa for hospitality.  }
%\newpage 

\textit{The third author is grateful to the Ministry of Education and Science of Ukraine for the support in completing the work `Artificial porous materials as a basis for creating the novel biosensors'. }

\newpage

{\footnotesize

%\vsk
O. Boyko$^1$, O. Martynyuk$^2$, V. Pivovarchik$^3$

$^{1,2,3}$ South Ukrainian National Pedagogical University, Staroportofrankovskaya str. 26, Odesa, Ukraine, 65020% Author's institution address, e-mail address.
 
 $^1$  boykohelga@gmail.com
 
 $^2$ martynyukolga@gmail.com
 
 $^3$ vpivovarchik@gmail.com - the corresponding author
 
%\vs

\received{?} %(to fill in by the Editors)

%%\revised{?} % (to fill in by the Editors)
\end{document}